\documentclass[]{aa}  
\usepackage{graphicx}
\usepackage{txfonts}

\usepackage[utf8]{inputenc} 
\usepackage[T1]{fontenc} 
\usepackage{lmodern} 
\usepackage{amsmath,amsfonts,amssymb} 
\usepackage{graphicx} 
\usepackage[colorlinks=true, allcolors=black]{hyperref} 

\newcommand{\revision}[1]{#1} 
\newcommand{\revisionb}[1]{#1} 
\newcommand{\LEt}[1]{} 
\newcommand{\revisionc}[1]{#1} 

\newcommand{\dd}{\; \text{d}} 
\newcommand{\e}{\text{e}} 
\newcommand{\TF}{\mathcal{F}} 
\newcommand{\ii}{\text{i}} 
\newcommand{\sinc}{\text{sinc}} 
\def\Xint#1{\mathchoice  
   {\XXint\displaystyle\textstyle{#1}}%
   {\XXint\textstyle\scriptstyle{#1}}%
   {\XXint\scriptstyle\scriptscriptstyle{#1}}%
   {\XXint\scriptscriptstyle\scriptscriptstyle{#1}}%
   \!\int}
\def\XXint#1#2#3{{\setbox0=\hbox{$#1{#2#3}{\int}$}
     \vcenter{\hbox{$#2#3$}}\kern-.5\wd0}}

\def\dashint{\Xint-}

\newcommand{\psiu}{\psi_\text{up}} 
\newcommand{\phiu}{\phi_\text{up}} 
\newcommand{\xiu}{\xi_\text{up}} 
\newcommand{\Pu}{P_\text{up}} 

\newcommand{\psid}{\psi_\text{do}} 
\newcommand{\phid}{\phi_\text{do}} 
\newcommand{\xid}{\xi_\text{do}} 
\newcommand{\Pd}{P_\text{do}} 

\newcommand{\M}{\mathcal{M}} 

\usepackage{schemabloc}
\usepackage{multicol}

\begin{document}

\title{Experimental validation of joint phase and amplitude wave-front sensing with coronagraphic phase diversity for high-contrast imaging.}
\titlerunning{Complex wavefront sensing with coronagraphic phase diversity}


   \author{O. Herscovici-Schiller \inst{1}
          \and
          L. M. Mugnier \inst{1}
          \and
          P. Baudoz \inst{2}
          \and
          R. Galicher \inst{2}
          \and
          J.-F. Sauvage \inst{1,3}
          \and
          B. Paul \inst{1}\thanks{Baptiste Paul is now with Thales Alenia Space}
          }

   \institute{ONERA -- The French Aerospace Lab\\
           F-92322 Châtillon, France\\
              \email{olivier.herscovici@onera.fr}
         \and
            LESIA, CNRS, Observatoire de Paris, Université Paris Diderot, Université Pierre et Marie Curie\\
       5 place Jules Janssen, F-92190 Meudon, France
       \and 
       Laboratoire d’Astrophysique de Marseille UMR 7326, Aix-Marseille Université, CNRS\\
             F-13388 Marseille, France
             }

   \date{7 December 2017}

  \abstract
   {The next generation of space-borne instruments dedicated to the direct detection of exoplanets requires unprecedented levels of wavefront control precision. Coronagraphic wavefront sensing techniques for these instruments must measure both the phase and amplitude of the optical aberrations using the scientific camera as a wavefront sensor.}
   {In this paper, we develop an extension of coronagraphic phase diversity to the estimation of the complex electric field, that is, the joint estimation of phase and amplitude. }
   {We introduced the formalism for complex coronagraphic phase diversity. We have demonstrated experimentally on the \textit{Très Haute Dynamique} testbed at the Observatoire de Paris that it is possible to reconstruct phase and amplitude aberrations with a subnanometric precision using coronagraphic phase
diversity. Finally, we have performed the first comparison between the  complex wavefront estimated using coronagraphic phase diversity (which relies on time-modulation of the speckle pattern) and the one reconstructed by the self-coherent camera (which relies on the spatial modulation of the speckle pattern).}
   {We demonstrate that coronagraphic phase diversity retrieves complex wavefront with subnanometric precision with a good agreement with the reconstruction performed using the self-coherent camera.}
   {This result paves the way to coronagraphic phase diversity as a coronagraphic wave-front sensor candidate for very high contrast space missions.
}


   \maketitle
%
\section{Introduction}
\label{sec:intro}  
One main science goal of future large space telescopes such as the Large UV/Optical/Infra\-red Surveyor (LUVOIR) or the Habitable Exoplanet Imaging Mission (HabEx) is exoplanet imaging and characterization. 
Direct imaging of exo-Earths represents a challenge on the instrumental point a view. \revisionc{A coronagraph (or an instrument that serves the same purpose, such as a starshade)} is needed to address the immense contrast between a Earth-like planet and its star: for example, the flux ratio between Earth and Sun is about $10^{-10}$ in the near infra-red. Besides, any aberration in the optical system causes light leakage in the coronagraph, which in turn generates speckles in the scientific images, thus limiting the detection level. Consequently, optical aberrations must be measured and corrected in order to avoid any false detection or biased characterization. To do so, both phase and amplitude aberrations must be measured and compensated. Moreover, the measurement must be performed from the science image to avoid \revisionc{non-common} path aberrations between the wavefront sensor and the science camera. In this article, we describe the extension of COFFEE, the coronagraphic phase diversity, to the estimation of both the phase and the amplitude defects of the light beam that propagates in a coronagraphic system. In addition, we demonstrate this capacity on experimental high contrast images. 

In Section \ref{sec:formalism}, we present the formalism of COFFEE, extended to take into account amplitude aberrations in the estimation process. Then, we present the experimental validation of this technique. In Section \ref{sec:calibrations}, we present the THD2 (\textit{Banc Très Haute Dynamique} version 2) experimental testbed (see \cite{thd2}), which reaches very high contrasts and allows for accurate phase and amplitude aberration control. We also explain the protocol of the experiment. In Section \ref{sec:ampl}, we present results on the estimation of a wave-front that is dominated by amplitude aberrations. Finally, in Section \ref{sec:onde}, we present the results of the retrieval of a wavefront containing both phase and amplitude aberrations. 

\section{Extension of COFFEE to amplitude estimation: formalism}
\label{sec:formalism}
\subsection{Model of image formation}
In this section, we describe how coronagraphic phase diversity can estimate amplitude aberrations.
Phase diversity (\cite{Gonsalves-a-82, Mugnier-l-06a}) relies on a model of image formation.
We have modeled an image of a point source whose flux is \(\alpha\), in the presence of a constant background \(\beta\), taken with a coronagraphic optical system whose response to an on-axis source is \(h_\text{c}\), and a detector whose impulse response is \(h_\text{det}\) as
\begin{equation}
\label{eq:modele_image}
\mathbf{I}(x,y) = \alpha\times [h_\text{det}\star\ h_\text{c}](x,y) + \beta + n(x,y),
\end{equation}
where \(n\) is the noise in the image, a subject we return to at the end of this subsection. We note that we consider the possibility that \(\alpha\) and \(\beta\) might change from one image to another.

Let us first detail the response of the optical system, \(h_\text{c}\). We rely on Fourier optics to describe the propagation of light in the system. In order to keep the same orientation for all the planes, we describe propagation from a pupil plane to a focal plane by an inverse Fourier transform, and propagation from a focal plane to a pupil plane by a direct Fourier transform, as in \cite{foo2005ol} or \cite{Herscovici-a-17}. The relevant parameters of the optical system are the (upstream) entrance pupil \(\Pu\), the (downstream) pupil of the Lyot stop \(\Pd\), and the focal-plane mask of the coronagraph  \(\mathcal{M}\). The parameters that we sought to retrieve are the complex aberration fields including phase and amplitude aberrations. We call \(\psiu\) the complex aberration field upstream of the coronagraph, and \(\psid\) the complex aberration field downstream of the coronagraph.

A natural and usual expression for a complex aberration field of amplitude \(A\) and phase \(\phi\) is \(\psi=A\e^{\ii\phi}\). However, in such a form, the amplitude aberration and the phase aberration play extremely asymmetrical roles. We wanted to avoid such an asymmetry because it might cause numeric difficulties while retrieving \(\psi\). Indeed, in this form, the gradient of the complex aberration field with respect to phase and amplitude are 
\[\dfrac{\partial A\e^{\ii\phi}}{\partial \phi} = \ii A\e^{\ii\phi} \text{ and } \dfrac{\partial A\e^{\ii\phi}}{\partial A} = \e^{\ii\phi} , \]
which are likely not to be of the same order of magnitude, resulting in numerical convergence problems of minimizers. On the contrary, if the complex field is represented by two parameters that play symmetric roles, this difficulty is avoided (\cite{Vedrenne-a-14}). Here, we chose to represent the complex fields by introducing the log-amplitude \(\xi = \log(A) \), resulting in \(\psi=\exp(\ii\phi+\xi)\). With these conventions, the coronagraphic intensity distribution for an on-axis source is written\LEt{or 'gives'?} 
\begin{align}
& h_\text{c}[\phiu,\xiu,\phid,\xid] = \\
& \left|\TF^{-1}\left\lbrace \Pd \e^{\ii\phid+\xid} \times  \TF \left[ \M \times \TF^{-1}(\Pu\e^{\ii\phiu+\xiu}) \right] \right\rbrace \right|^2. \nonumber
\label{eq:hc}
\end{align}
Hereafter, we will suppose that \(\xid=0\), that is to say we suppose that there is no downstream amplitude aberration, or at least that the downstream amplitude is known and taken into account in \(\Pd\).

As for the noise \(n\), it is the result of two main contributions. The first one source is the detector read-out noise, which is classically modeled as a spatially homogeneous random white Gaussian process for a charge-coupled device detector. The calibration of the detector read-out noise can be performed prior to the experiment. The second contribution is photon noise. It is modeled as a random Poisson process, and can be well-approximated by a nonhomogeneous Gaussian white noise. Since \(n\) is the sum of two Gaussian white noises, it is a nonhomogeneous Gaussian white noise.

\subsection{COFFEE, a Bayesian maximum a posteriori estimator}
COFFEE is an extension of phase diversity described in \cite{Paul-a-13b}. It relies on the same maximum \LEt{please remove the italics.}a posteriori approach: it retrieves the unknown parameters by fitting an image model to experimental data, using knowledge on the statistics of the noise and \LEt{please remove the italics.}a priori information on the unknown parameters.

It requires several images with a known introduced phase difference between them, in order to be able to determine the aberrations of the optical system unambiguously. While the classical phase diversity technique generally uses only two images, at least three images are necessary to retrieve both phase and amplitude aberrations in practice. This can be understood by the fact that more data are needed to reconstruct three maps (\(\phiu\), \(\phid\) and \(\xiu\)) than two. This has been confirmed by similar works in a different context (\cite{Vedrenne-a-14}). 
Consequently, the experimental data that we take and process will always contain at least three images, differing only by a known phase diversity. Moreover, in the case of coronagraphic phase diversity, the diversity phases must be introduced upstream of the coronagraph.

We denote by \(\mathbf{I}_k\) the image taken with an introduced phase diversity \(\phi_{\text{div}, k}\). The index \(k\) refers to the choice of diversity phase. We have taken the convention that \(k=0\) always denotes an image with no diversity.
We denote by \(\mathbf{I}_k(x,y)\) the pixel of coordinates \((x,y)\) of the image \(\mathbf{I}_k\). For example, in the experimental part of this paper from Sections~\ref{sec:calibrations} to~\ref{sec:onde}, we use three different images --- the index \(k\) ranges from 0 to 2; and the data are 360$\times$360-pixel images --- indexes \(x\) and \(y\) go from 1 to 360. 

Considering the form of our image model (Eq.~\ref{eq:modele_image})
for each \(\mathbf{I}_k\), the noiseless image model taken with diversity phase  \(\phi_{\text{div},k}\) is
\begin{align}
&\mathbf{M}[\alpha,\beta,\phiu,\xiu,\phid](k,x,y) = \\ &\alpha(k)\times \left\{h_\text{det}\star\ h_\text{c}[\phiu+\phi_{\text{div},k},\xiu,\phid]\right\}(x,y) + \beta(k).\nonumber
\end{align}

Since \(n\) is a Gaussian white noise, the unknown parameters are the ones that minimize the following penalized least-squares criterion (\cite{idier08}):
\begin{align}
\label{eq:principe_COFFEE}
& J({\phiu,\xiu,\phid}) = \\
& \hspace{1cm} \sum_{(k,x,y) }\left| \dfrac{\mathbf{I}_k(x,y) - \mathbf{M}[\phiu,\xiu,\phid](k,x,y)}{\sigma_n(k,x,y)} \right|^2 \nonumber \\
& \hspace{1cm} + \mathcal{R}(\phiu,\xiu,\phid). \nonumber
\end{align}
Here,  \(\sigma_n\) is the map of standard deviation of the noise~\(n\) and \( \mathcal{R} \) is a regularization term.
 
\subsection{Regularization}
The regularization term, \(\mathcal{R}\), represents \LEt{please remove the italics.}a priori information on the unknowns. These unknowns are numerous: \(\alpha\), \(\beta\), \(\phiu\), \(\xiu\), and \(\phid\). Both \(\alpha\) and \(\beta\) are scalars for each image, and there exists an analytic solution for them. The other unknowns are three maps of typically \(40 \times 40\) elements -- \(40\times 40\) because we aim to reconstruct aberration maps at a resolution better than the number of actuator on a \(40 \times 40\) deformable mirror, using images sampled typically at the Shannon-Nyquist limit. There is no analytic solution for this problem. We used the VMLM-B method of \cite{Thiebaut-p-02} to solve it numerically. Since the problem is not heavily over-determined, with typically \(3\times 40 \times 40 + 3 \times 2 = 4806 \) unknowns versus typically \(3\times 80 \times 80 = 19200\) noisy and partially redundant data points, the stability of the reconstruction can only be obtained by means of regularization.

We assumed that the energy spectrum densities of \(\phiu\), \(\xiu\) and \(\phid\) decrease as \( 1/f^2 \), where \( f\) is the norm of the spatial frequency, which is a classic (\cite{Church88}) and realistic assumption (\cite{Hugot12}). Hence, \(\mathcal{R}\) \revisionc{is written}\LEt{Please check I have retained your intended meaning.}
\begin{equation}
\label{eq:regul_basique}
\mathcal{R}(\phiu,\xiu,\phid) = \dfrac{\|\nabla \phiu \|^2}{2\sigma^2_{\nabla \phiu}} + \dfrac{\|\nabla \xiu \|^2}{2\sigma^2_{\nabla \xiu}} + \dfrac{\|\nabla \phid \|^2}{2\sigma^2_{\nabla \phid}} .
\end{equation}
The variances of \(\nabla\phiu\), \(\nabla\xiu\), and \(\nabla\phid\) are \revisionc{denoted by} \(\sigma^2_{\nabla \phiu}\), \(\sigma^2_{\nabla \xiu}\), and \(\sigma^2_{\nabla
\phid}\) respectively, \revision{where \(\nabla\) is the gradient with respect to the Cartesian space coordinates in the pupil plane}. They are computed analytically from \LEt{please remove the italics.}a priori information on the variances of \(\phiu\), \(\xiu\), and \(\phid\), as described in \cite{Paul-a-13b}.

\subsection{Distinctive features of COFFEE}
\revision{
Fundamentally, COFFEE relies on a physical, nonlinear model of image formation. This is its most distinctive feature, since other methods \revisionb{such as speckle nulling (\cite{trauger2004}), the self-coherent camera (\cite{galicher2008wavefront}), the electric field conjugation (\cite{Riggs_2016}) or the wavefront sensing with random DM probes (\cite{pluzhnik2017}) rely on} a linear or linearized model of the relationship between the aberrations and the image. This yields specific advantages and disadvantages.}

\revision{
\revisionb{On the one hand, among} the advantages of the COFFEE approach \revisionb{is the fact that} the quality of the COFFEE reconstruction is not affected by the measured wave-front aberrations not being very small compared \revisionb{to} the observation wavelength. COFFEE is therefore not limited to the estimation of small phase aberrations. This point is particularly helpful when initiating the \revisionb{Dark Hole (\cite{malbet1995}) process} (with possibly large static aberrations). Moreover, COFFEE does not require \revisionb{updating of a} calibration matrix during the Dark Hole process, contrarily to techniques \revisionb{such} as electric field conjugation.
Other advantages are that COFFEE needs no hardware modification to the coronagraphic system; and COFFEE is in theory not restricted to a monochromatic wavelength, even if \revisionb{computing cost would be higher if a wide-band image were to} be modeled.}

\revision{
On the other hand, the main current limitation of COFFEE is that it needs an accurate model of the instrument -- essentially in terms of image sampling, characteristics of the coronagraph, pupil geometry and wavelength -- in order to make precise estimates. Any model error results in error in the estimates. Also, we note that it currently takes about a minute to obtain a COFFEE estimate.
}Now that we have detailed the formalism of the method, we move on to its experimental validation\LEt{single-sentence paragraphs are not allowed.}.   

\section{Strategy of experimental validation}
\label{sec:calibrations}
\subsection{The THD2 testbed}
\begin{figure}[!ht]
\centering
\includegraphics[width=\linewidth]{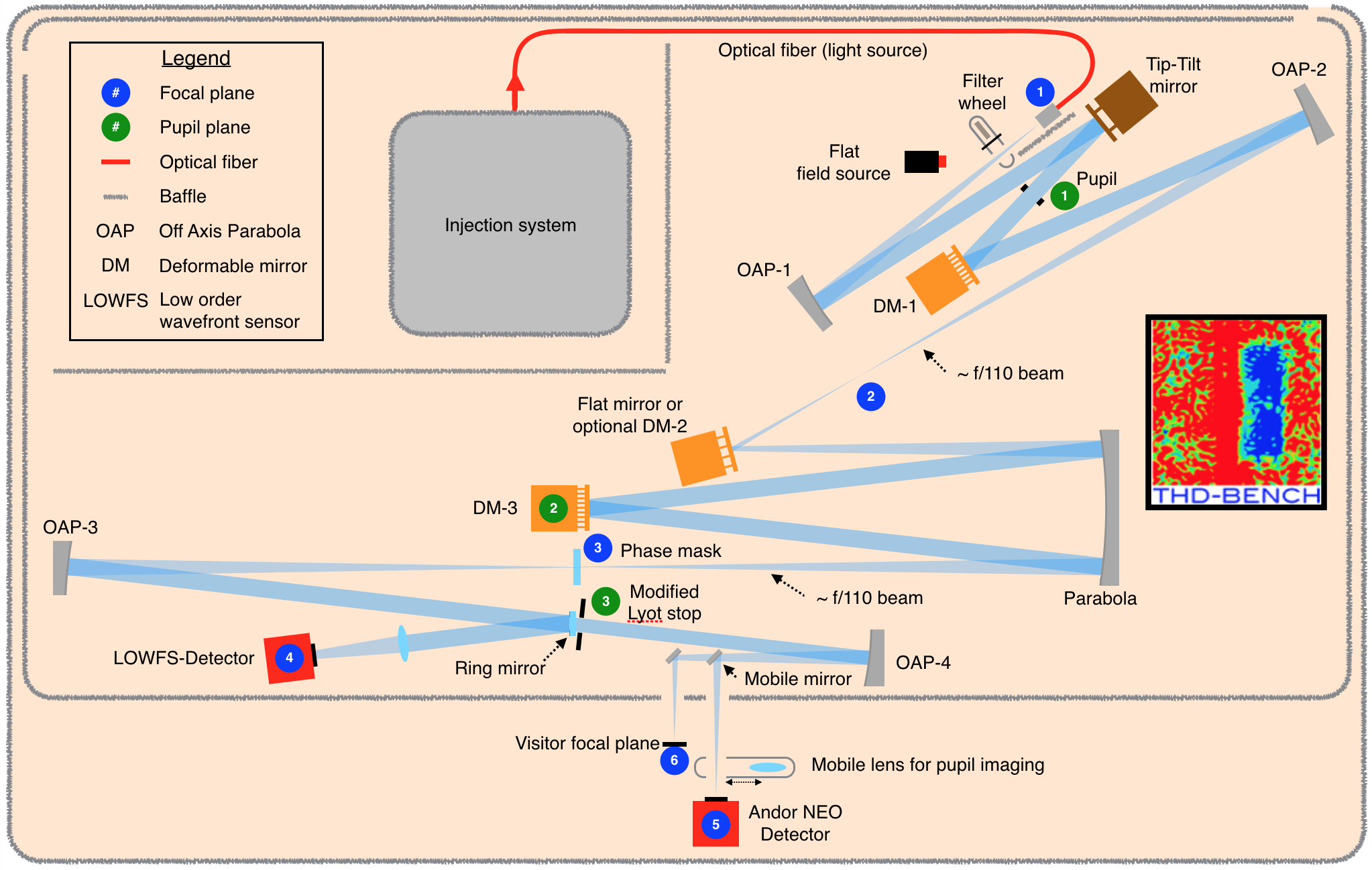}
\caption{Schematic representation of the THD2 bench.}
\label{fig:thd}
\end{figure}

We validate the joint phase and amplitude retrieval on the THD2 bench (\textit{Très Haute Dynamique} version~2). This very high contrast testbed at LÉSIA (Observatoire de Paris) is described in detail by \cite{thd2}, and represented in Fig.~\ref{fig:thd}. Its very high quality enables one to routinely reach contrasts down to \(2\times 10^{-8}\). 

For the sake of our experiments; let us mention here that it is equipped with:
\begin{itemize}
\item a monochromatic light source of wavelength 783.25 nm that feeds the bench through a single mode fiber, in focal plane 1, see Fig~\ref{fig:thd};
\item a photometer for the precise normalization of the light flux, integrated in the injection system; 
\item an out-of-pupil deformable mirror, DM--1, hereafter called “amplitude deformable mirror”;
\item a pupil-plane deformable mirror, DM--3, hereafter called “phase deformable mirror”;
\item a four-quadrant phase mask coronagraph (\cite{rouan2000}), with the focal mask in focal plane 3 and the Lyot stop in pupil plane 3;
\item a CMOS camera in focal plane 6, which can also be used for pupil plane imaging thanks to a movable lens.
\end{itemize}

\revision{We note that, although we call DM--1 the “amplitude mirror” for the sake of \revisionb{simplicity}, a deformation of this mirror introduces both amplitude and phase for most spatial frequencies. }

\subsection{Model calibration}
All the parameters of the model must be calibrated precisely. Any calibration error has an impact on the quality of the reconstruction of the aberrations. Here, we detail the calibrations performed on the THD2.
\paragraph{Calibration of the pixel response}
The pixel transfer function is simply modeled here by a top hat window function, parameterized by the size of the pixel. Calibrating the pixel size corresponds to calibrating the sampling factor of the detector with respect to the size of the diffraction.
Numerically, it determines \(h_\text{det}\) in Eq.~(\ref{eq:hc}).
To determine it, we record a noncoronagraphic image, \(i_\text{nc}\). The corresponding transfer function, \(\left|\TF(i_\text{nc})\right|^2\), reaches zero at a cut-off frequency \(f_\text{cut}\) and is sampled up to a maximum frequency \(f_\text{Nyquist}\). The sampling factor \(s\) is simply the ratio
\begin{equation}
s = 2\frac{f_\text{Nyquist}}{f_\text{cut}}
.\end{equation}

\begin{figure}[!ht]
\centering
\begin{tabular}{cc}
\includegraphics[height=0.35\linewidth]{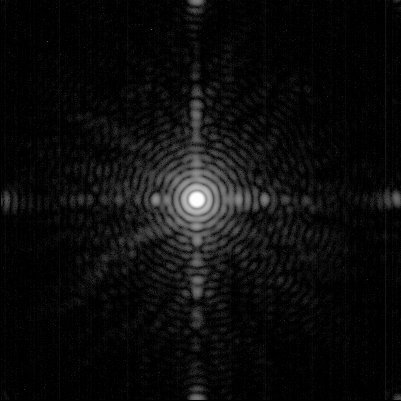} & \includegraphics[height=0.4\linewidth]{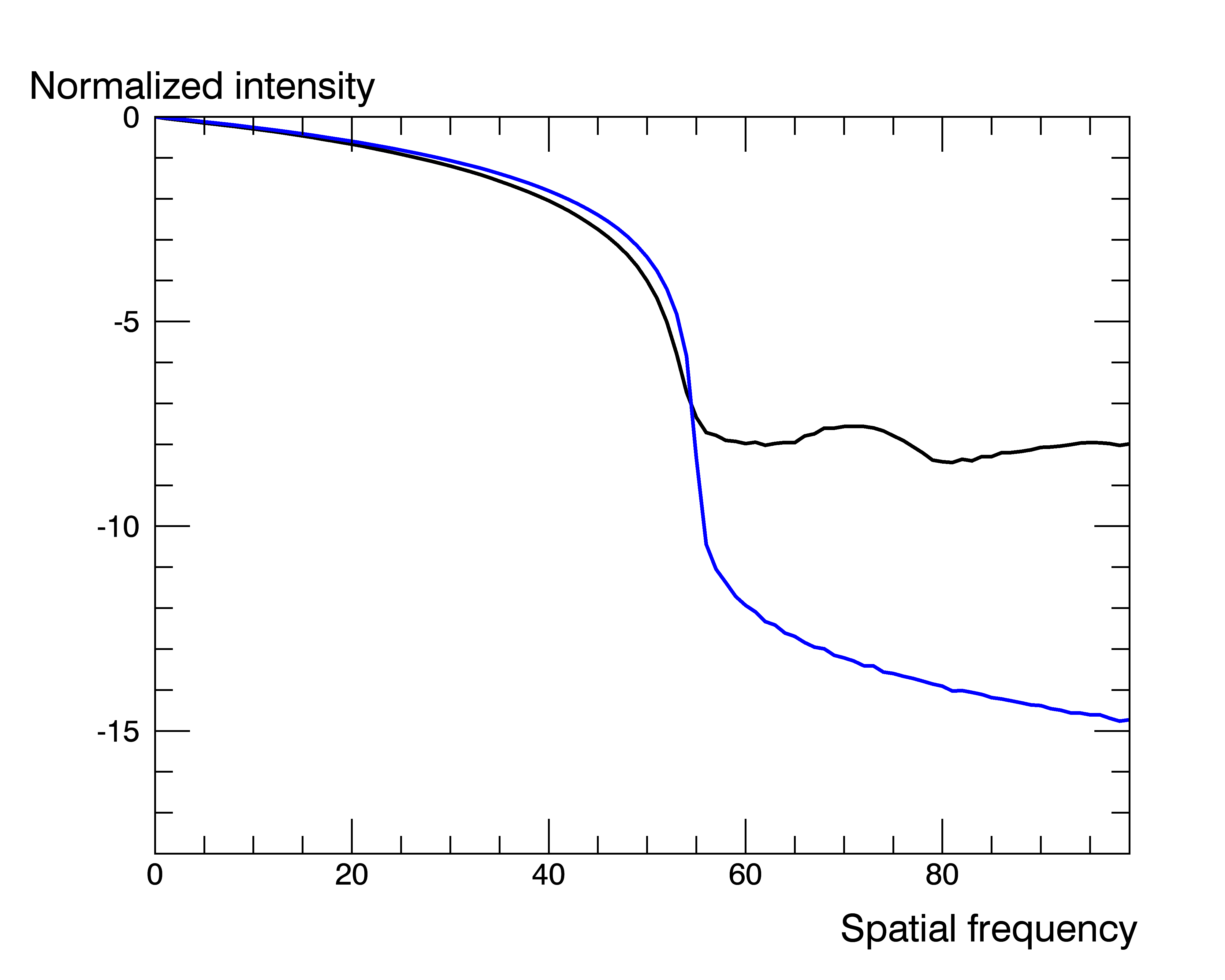}
\end{tabular}
\caption{Left: Noncoronagraphic PSF, in logarithmic scale. Right: Corresponding MTF (black) and MTF of an Airy pattern with same cut frequency (blue), in logarithmic scale. The \(x\)-axis has been cut at half the Nyquist frequency (200).}
\label{fig:non-corono}
\end{figure}

The experimental \(400\times400\) image and the corresponding modulation transfer function are displayed in Fig.~\ref{fig:non-corono}. In our case, \(f_\text{cut} = 56 \pm 0.3\) and \(f_\text{Nyquist}=200\), which yields a sampling factor \(s=7.14 \pm 0.04 \). The cross shape, with residuals on the axes, is due to the use of a four-quadrant phase mask coronagraph: this noncoronagraphic image has been obtained after the deformable mirrors were flattened using the self-coherent camera (\cite{Mazoyer-a-13}). Since the four-quadrant phase mask is indifferent to aberrations that create speckles located on the axes, the correction cannot be performed on the axes; when the four-quadrant phase mask is removed, the on-axes speckles are no longer filtered.

\paragraph{Lyot ratio} The Lyot ratio is defined as the ratio between the radius of the Lyot stop and the radius of the entrance pupil. In the terms of Eq.~(\ref{eq:hc}), it determines the radius of \(\Pd\) with respect to that of \(\Pu\). Here, we used a configuration of the THD2 where the diameter of \(\Pd\) is  \(6.5\)\,mm and the diameter of \(\Pu\) is \(8.23\)\,mm.

\paragraph{Detector noise and bias} According to the specification of our \revision{detector}, the standard deviation of its read-out noise is one electron. This is consistent with the value found by averaging the root mean square on the pixels of 6\,400 images acquired with the light source switched off (“background frames”). Each time a series of images is taken, a corresponding series of background frames is taken, and the median is subtracted from the science image in order to compensate for the bias of the detector. 

\paragraph{Diversity phases} For the COFFEE technique, as for any flavor of phase diversity, the diversity phases \(\phi_\text{div}\) that we introduce must be absolutely calibrated. 
Any imprecision on the knowledge of \(\phi_\text{div}\) will have direct repercussions of the same order of magnitude on the reconstructed parameters. The simplest way to introduce \(\phi_\text{div}\) is to use the phase deformable mirror, DM3. However, until now, the minimization of speckles intensity on the THD2 experiment was only done using the Self-Coherent Camera as a focal plane wave-front sensor in closed loop. So neither the estimated wavefront nor the DM response to voltages (\cite{DM3}) required absolute calibration. In order to use COFFEE on the THD2 bench, we calibrated the DM3 response to a given set of control voltages the response of this mirror (a 1024-actuator MEMS Boston Micromachine) to a given set of control voltages.
We will detail the procedure for the first diversity map, \(\phi_{\text{div},k=1}\). The principle is the same for \(\phi_{\text{div},k=2}\); and \mbox{\(\phi_{\text{div},k=0}\)}, is taken equal to 0 (we use a focused image). 

\revision{We chose defocus as a diversity \revisionb{because it is the most used one} in noncoronagraphic phase diversity. More precisely, the focus shape that can be achieved by a 32x32 deformable mirror is quite good, but not a \revisionb{pure} defocus. Apart from the amplitude error, there is also a small shape error. As we need to know precisely the diversity phase shape, we \revisionb{first} performed a calibration of the latter. 
} In order to calibrate the diversity map \(\phi_{\text{div},k=1}\) using our set-up, the easiest way is to use the noncoronagraphic phase diversity technique itself. The method is quite straightforward.\LEt{A\&A avoids lists where possible, please consider re-formatting this as one or mroe paragraphs, possibley with a 'firstly... secondly...' structure.} 
Firstly, we applied a command to the phase deformable mirror that produces \(\phi_{\text{div},k=1}\), which is the phase that we wanted to calibrate. 
Secondly, we recorded an image \(i_\text{calib, 0}\). 
Thirdly, we mechanically moved the detector \revision{\(12.70\pm 0.02\) mm away from its nominal} position at the focus.
This action on the position of the detector is optically equivalent to introducing a pure defocus whose amplitude is given by Eq.~(12) of \cite{Blanc-a-03a}. 
Fourthly, we acquired a second image \(i_\text{calib, 1}\) at this position, before returning the detector to its original position. 
Fifthly, we used \(i_\text{calib, 0}\) and \(i_\text{calib, 1}\) as input data for phase diversity, which estimates the sum of \(\phi_{\text{div},k=1}\) and \(\phiu^0\), with \(\phiu^0\) the unavoidable static aberration that exists on the bench. 

To calibrate \(\phiu^0\), we repeated the same procedure using \(\phi_{\text{div},k=0} = 0\) instead of \(\phi_{\text{div},k=1}\). 
Finally, we obtain (by subtraction) \(\phi_{\text{div},k=1}\). The same complete procedure yields \(\phi_{\text{div},k=2}\). Both results\LEt{Please check I have retained your intended meaning.} are
quite different from a defocus\revision{; their structure reflects the imperfection of the deformable mirror}. \revision{A Zernike decomposition of \(\phi_{\text{div},k=1}\) shows that defocus accounts for only about 80\% of the total phase variance of \(\phi_{\text{div},k=1}\), whose root mean square value of \(\phi_{\text{div},k=1}\) is 19 nm.} The same applies to \(\phi_{\text{div},k=2}\), whose root mean square value is 29 nm. \revision{\revisionb{Using Eq. (12) of \cite{Blanc-a-03a}}, the imprecision on these measurements due to the propagation of the error on the displacement of the detector is 0.2 nm.} 
\begin{figure}[!ht]
\centering
\includegraphics[width=\linewidth]{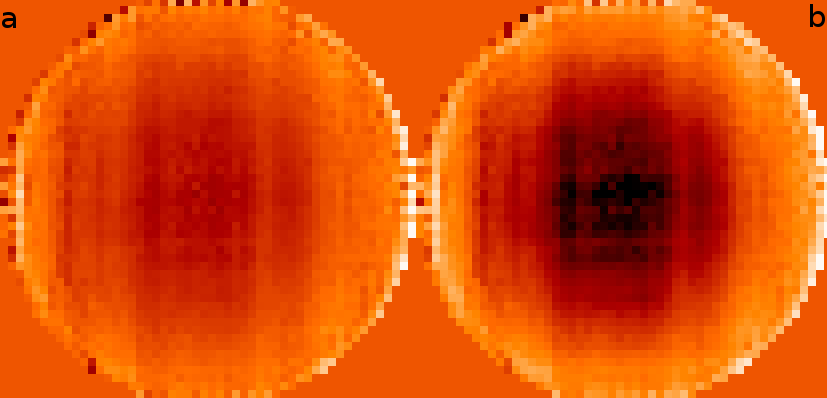} 
\caption{a. \(\phi_{\text{div},(k=1)}\) (left). b. \(\phi_{\text{div},(k=2)}\) (right).}
\label{fig:phase_div}
\end{figure}

\revisionc{The model of the experimental set-up is now calibrated. The next step is to look more closely at the regularization.}\LEt{single sentence paragraphs are not allowed.}

\subsection{Regularization strategy for the four-quadrant phase mask coronagraph}
A feature of the four-quadrant phase mask coronagraph is that it is insensitive to some particular phase modes. Indeed, let us denote by \(\phi_0\) a symmetric phase such that \(\TF(\phi_0)\) be significantly different from zero only on the axes. Then the model of the corresponding image is the same as the one obtained with a perfectly flat input wave-front: \(h_\text{c}(\phiu\!=\!\phi_0) = h_\text{c}(\phiu\!=\!0) \) --- see Appendix \ref{ann:modes_non-vus} for the derivation. Consequently, when analyzing images taken with a four-quadrant phase mask such as the one we used here, COFFEE is insensitive to any linear combination of such modes. From an inverse problem point of view, this means that the forward model is noninjective, which implies that the reconstruction needs to be regularized. This is analogous to a classic problem in adaptive optics with a Shack-Hartmann wavefront sensor: just as the waffle mode is unseen\LEt{not seen?.} by the Shack-Hartmann wavefront sensor and must be filtered out of the control in order not to saturate the deformable mirror, here the “cross” modes are unseen\LEt{not seen?.} by the four-quadrant phase mask and must be filtered in order not to saturate the reconstruction of our focal-plane wavefront sensor. Indeed, we have checked that if this problem is not dealt with, the estimates always go to unrealistic root mean square values, and a Fourier analysis of the estimates shows significant values only on the axes.
To address this issue, we added another regularization term to the usual one expressed by Eq.~(\ref{eq:regul_basique}). This regularization must prevent any term of the form \(\phi_0\) to become dominant in the estimation of \(\phiu\). We chose the following quadratic, hence convex and differentiable, functional:
\begin{equation}
\label{eq:regul_4Q}
\mathcal{R}_\text{FQPM}(\phiu) = \dfrac{\eta}{2\sigma_\phi ^2}\left\| \chi \times \TF^{-1}[\phiu] \right\|^2;
\end{equation}
where the hyperparameter \(\eta\) is typically on the order of ten, \(\sigma_\phi\) is the a priori\LEt{please remove the italics.} information on the standard deviation of \(\phiu\), and \(\chi\) is a weighting function equal to 1 in an area of \(1 \lambda/D\) around the axes of the four-quadrant phase mask and zero elsewhere. 
The gradient of this term is useful for numeric minimization. Its expression is simply
\begin{equation}
\dfrac{\partial \mathcal{R}_\text{FQPM}(\phiu)}{\partial \phiu} = \dfrac{\eta}{\sigma_\phi ^2}\times \TF\left\{\chi^2\times\TF^{-1}\left[\phiu\right]\right\}.
\end{equation}

\subsection{Wavefront measurement strategy: differential measurements}

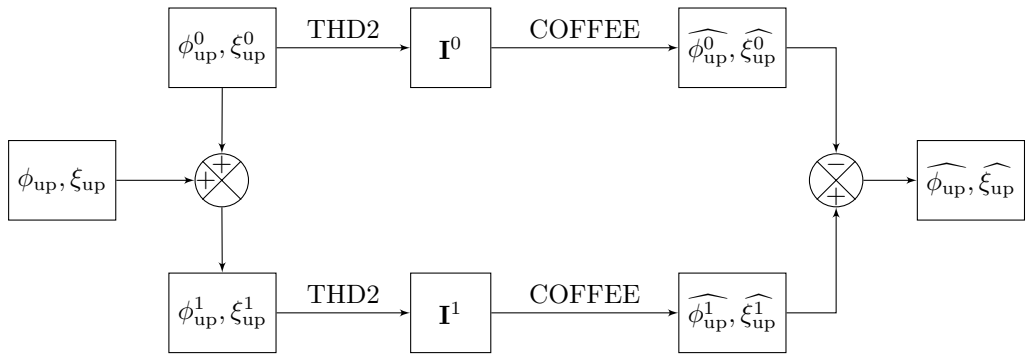
\begin{figure*}[!ht]
\centering
{\normalsize
\begin{tikzpicture}
\sbEntree{E1}
\sbBloc[1]{param}{\(\phiu^1,\xiu^1\)}{E1} 
\sbBloc[5]{im}{\(\mathbf{I}^1\)}{param} 
\sbRelier[THD2]{param}{im}
\sbBloc[7]{estimee}{\(\widehat{\phiu^1},\widehat{\xiu^1}\)}{im} 
\sbRelier[COFFEE]{im}{estimee}

\sbDecaleNoeudy[-10]{E1}{E2}
\sbBloc[1]{paramref}{\(\phiu^0,\xiu^0\)}{E2} 
\sbBloc[5]{imref}{\(\mathbf{I}^0\)}{paramref} 
\sbRelier[THD2]{paramref}{imref}
\sbBloc[7]{estimeeref}{\(\widehat{\phiu^0},\widehat{\xiu^0}\)}{imref} 
\sbRelier[COFFEE]{imref}{estimeeref}

\sbDecaleNoeudy[-5]{E1}{E3}
\sbSumh[3]{somme}{E3}
\sbBloc[-5]{inconnue}{\(\phiu,\xiu\)}{E3} 
\sbRelier{paramref}{somme}
\sbRelier{inconnue}{somme}
\sbRelier{somme}{param}

\sbCompSum[26]{comp}{E3}{-}{+}{}{}
\sbRelierxy{estimee}{comp}
\sbRelierxy{estimeeref}{comp}

\sbBloc[2]{final}{\(\widehat\phiu,\widehat\xiu\)}{comp}
\sbRelier{comp}{final}

\end{tikzpicture}}
\caption{Synthetic representation of the validation strategy.}
\label{fig:strategie}
\end{figure*}

A precise characterization of a wavefront sensor can only be done on a bench with a calibrated wavefront. On THD2, one term is unknown: the bench’s own amplitude and phase aberrations \(\phiu^0,\xiu^0,\phid^0\). Even if these aberrations are extremely small (leading to a \(10^{-8}\) contrast in intensity), they bias our estimation and must be calibrated. We used a classical method of differential measurements to remove the contribution of the bench own aberrations to the result: in order to compare COFFEE reconstructions to known aberrations, we compare differential COFFEE reconstructions to known differential aberrations. More precisely, we perform a COFFEE reconstruction \(\widehat{\phiu^0},\widehat{\xiu^0}\) of the aberrations \(\phiu^0,\xiu^0\) on the THD2 in its reference state, indicated by index 0. We then introduced a supplementary upstream aberration of known characteristics \(\phiu,\xiu\), an perform a COFFEE reconstruction \(\widehat{\phiu^1},\widehat{\xiu^1}\) of this aberration \(\phiu^1,\xiu^1,\phid^0\) = \((\phiu,\xiu,0) + (\phiu^0,\xiu^0,\phid^0)\). We computed the difference of the two reconstructions, and finally, we compared this difference with the known introduced aberrations \(\phiu, \xiu\). This process is presented schematically in Fig.~\ref{fig:strategie}. 

\subsection{Measurement of the reference wavefront}
\label{subsec:reference}
\begin{figure}[!ht]
\centering
\includegraphics[height=0.33\linewidth]{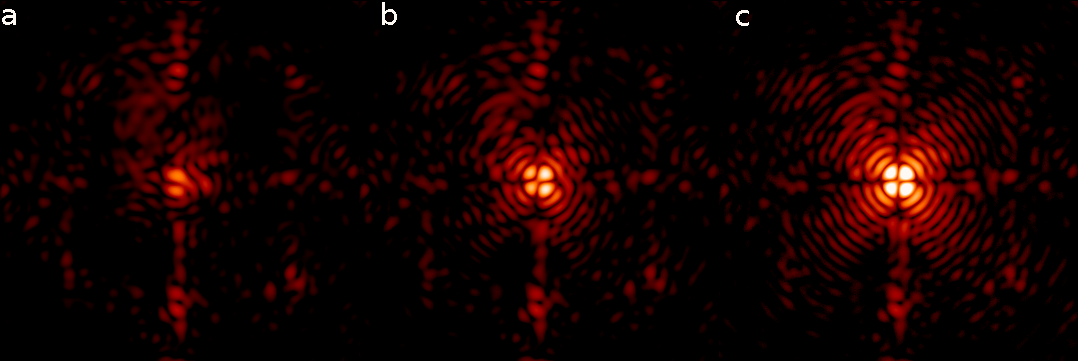}\includegraphics[height=0.33\linewidth]{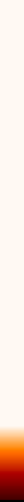}

\caption{a. Focused image \(\mathbf{I}^0_{k=0}\) (left). b. Diversity image \(\mathbf{I}^0_{k=1}\) (middle). c. Diversity image \(\mathbf{I}^0_{k=2}\) (right). The scale is an argument hyperbolic sine, with the same color scale for all the images.}
\label{fig:I0}
\end{figure}

Here we describe the operations corresponding to the top horizontal branch of Fig.~\ref{fig:strategie}.
The estimation of the reference wavefront is done as follows.
Reference wavefront controls were imposed on the phase mirror and on the amplitude mirror, generating phase \(\phiu^0\) and log-amplitude \(\xiu^0\). The corresponding data \(\mathbf{I}^0_{k=0}\) is acquired.
For the acquisition of the first diversity image \(\mathbf{I}^0_{k=1}\), a control voltage corresponding to the first diversity (Fig.~\ref{fig:phase_div}, left) is added to the phase mirror, so the phase becomes \(\phiu^0+\phi_{\text{div},k=1}\) and the amplitude is unchanged. In a similar fashion, for the acquisition of the second diversity image \(\mathbf{I}^0_{k=2}\), a control voltage corresponding to the second diversity (Fig.~\ref{fig:phase_div}, right) was added to the phase mirror, so the phase becomes \(\phiu^0+\phi_{\text{div},k=2}\). The images \(\mathbf{I}^0_{k=0}\), \(\mathbf{I}^0_{k=1}\) and \(\mathbf{I}^0_{k=2}\) are displayed in Fig.~\ref{fig:I0}.

\begin{figure}[!ht]
\centering
\includegraphics[width=\linewidth]{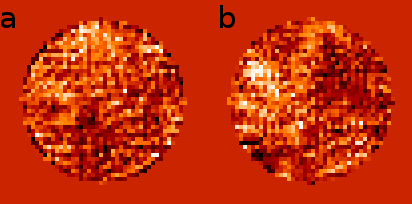}
\caption{a. \(\widehat{\phiu^0}\) (left). b. \(\widehat{\xiu^0}\) (right).}
\label{fig:estimee_ref}
\end{figure}

Using these images, we retrieved the COFFEE estimates of the reference phase \(\widehat{\phiu^0}\) and the reference log-amplitude \(\widehat{\xiu^0}\), which are displayed in Fig.~\ref{fig:estimee_ref}. The root mean square values of the reconstructed aberrations are \(\sigma(\widehat{\phiu^0}) = 3.4\) nm and \(\sigma(\widehat{\xiu^0}) = 2.9\) nm. The complete set of parameters used for the reconstruction is displayed in Table~\ref{tab:param_ref}. We note that the reconstruction is very robust, that is to say insensitive to the \LEt{please remove itlaics.}a priori values of the standard deviations of \(\phiu\), \(\xiu\) and \(\phid\). For example, the retrieved phase using the parameters in Table~\ref{tab:param_ref} has a correlation of \revision{0.999998} and a relative difference of \(4.7\times10^{-5}\) with the retrieved phase using \(\sigma(\phiu) = 30\)\ nm, \(\sigma(\xiu) = 25\)\ nm and \(\sigma(\phid) = 5\)\ nm. \revision{Also, the actual power spectrum density is different from a \(1/f^2\) power law, because the reference state is such that the deformable mirror DM3 partially corrects the  phase defects up to its maximum spatial frequency, beyond which it cannot perform \revisionb{any} correction.}

\begin{table}
\caption{Parameters of reconstruction for the reference wavefront\LEt{please remove itlaics form "a priori.}.}

\tiny
\centering
\begin{tabular}{l r}
\hline
Parameter & Value\\
\hline
Wavelength \(\lambda\) & 783.25\,nm \\
Data set & \(\mathbf{I}^0\) \\
Images size & \(360\times360\) pixels \\
Sampling factor & 7.14 \\
Lyot ratio & 0.759 \\
Diversity phases & \(\phi_{\text{div},(k=1)}\) and \(\phi_{\text{div},(k=2)}\) \\
Light flux on the photometer for \(\mathbf{I}^0_{k=0}\) & \(2.5\times10^{-6}W\)\\
Light flux on the photometer for \(\mathbf{I}^0_{k=1}\) & \(5.2\times10^{-7}W\)\\
Light flux on the photometer for \(\mathbf{I}^0_{k=2}\) & \(1.5\times10^{-7}W\)\\
Number of averaged frames for \(\mathbf{I}^0_{k=0}\) & 800 \\
Number of averaged frames for \(\mathbf{I}^0_{k=1}\) & 800 \\
Number of averaged frames for \(\mathbf{I}^0_{k=2}\) & 900 \\
Read-out noise standard deviation & 1 electron \\
Coronagraph type & 4-quadrant phase mask \\
A priori on \(\sigma_{\phiu^0}\) & 2\ nm\\
A priori on \(\sigma_{\phid^0}\) & 0.5\ nm\\
\textit{A priori} on \(\sigma_{\xiu^0}\) & 2\ nm\\
\hline
\end{tabular}

\label{tab:param_ref}
\end{table}
 
\section{Experimental retrieval of a known wavefront dominated by amplitude aberration}
\label{sec:ampl}

In this section, we generate a wavefront that is dominated by amplitude aberration.
 We do not change the command on the phase mirror, DM3.
 On the off-pupil amplitude mirror, DM1, we apply a sinusoidal aberration whose frequency \(\nu\) is chosen such that Talbot effect \revision{(\cite{Zhou2010})} converts the off-pupil phase map of DM1 into a pure amplitude aberration. \revisionb{Talbot effect or self-imaging appears when observing the Fresnel diffraction of a sinusoidal pattern at fraction or multiple distance of the Talbot length $z_T=2/(\nu^2\lambda)$. As shown in \cite{Zhou2010}, an exact image of a pure sinusoidal phase aberration will appear at distances $z = n z_T/2$ (with $n$ an integer). At distances  $z = (2n+1) z_T/4$, the field will be converted to pure sinusoidal amplitude aberration. Applying this last equation with $n=0$ to the THD2 bench, i.e. a distance DM1-pupil $z=269$ mm and a wavelength $783.25$ nm, the first sinusoidal frequency that will be completely converted to amplitude is equal to $\nu=1.54$ mm$^{-1}$ (period of 0.65 mm). We applied such a frequency avoiding the four quadrant transition direction by $22.5^{\circ}$.}
Taking again the notations of Fig.~\ref{fig:strategie}, this means that \(\phiu=0\), and we aim for \(\xiu(r) = C \sin(2\pi \nu\cdot r) \).

\begin{figure}[!ht]
\centering
\includegraphics[height=0.33\linewidth]{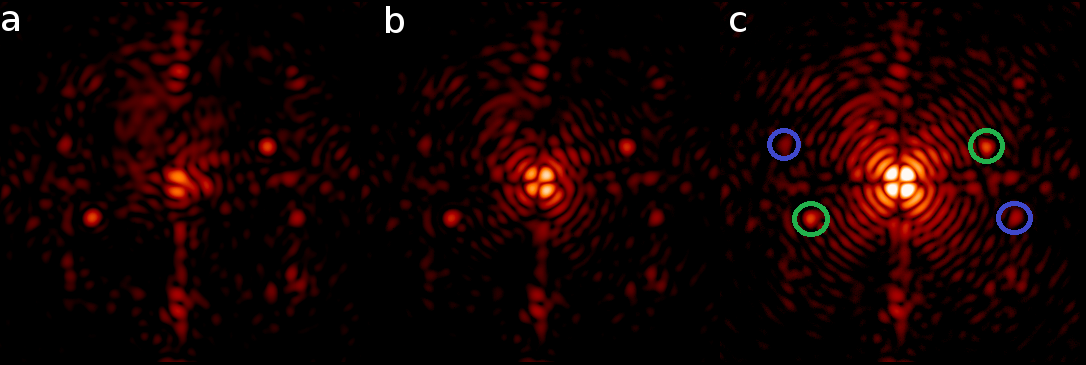}\includegraphics[height=0.33\linewidth]{colorbar}
\caption{a. Focused image \(\mathbf{I}^1_{k=0}\) (left). b. Diversity image \(\mathbf{I}^1_{k=1}\) (middle). c. Diversity image \(\mathbf{I}^1_{k=2}\) (right). The scale is an argument hyperbolic sine. On the rightmost image, the main pair of spots is enhanced by green circles, and the replica spots are enhanced by blue circles. They are clearly visible in all three images.}
\label{fig:I1}
\end{figure}

However, the \(32 \times 32\)-actuator DM1 cannot produce a continuous sinusoid but only an approximate sinusoid. Consequently, instead of generating only a pair of spots, as would be the case if the deformable mirror had an infinite number of actuators, the deformable mirror generates several pairs of spots. \revision{In Appendix \ref{ann:DM}, we describe a kind of “dual-aliasing” effect which explains that any continuous field (the electromagnetic field) that encounters a spatially discrete modulation exhibits unexpected resonances that in turn result in \revisionb{these} unexpected ghost spots.} Figure~\ref{fig:I1} displays the focused and the two diversity images taken with this amplitude aberrations. On the rightmost image of Fig.~\ref{fig:I1}, the green circles show the main pair of spots (the one that would be generated by a deformable mirror with infinitely many actuators) and the blue ones mark replica spots due to the discrete nature of the deformable mirror.
Fresnel propagation for the pair of secondary spots predicts a mixture of phase and amplitude in a pupil plane. 

The corresponding data set is acquired just like in the previous section, and displayed in Fig.~\ref{fig:I1}. \revisionc{We note the apparition of pairs of bright spots that are absent in} Fig.~\ref{fig:I0}\LEt{Please check I have retained your intended meaning.}. They are the manifestation of the periodic amplitude aberration. 

\begin{figure}[!ht]
\centering
\includegraphics[width=\linewidth]{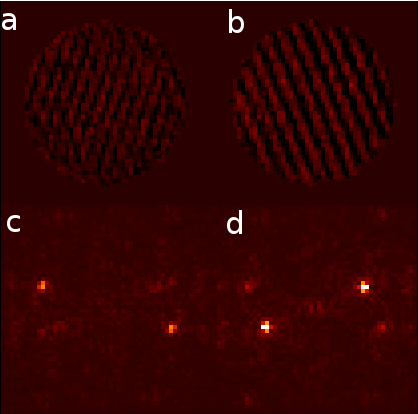}
\caption{a. Estimated phase (top left). b. Estimated log-amplitude (top right). The linear color bar extends from -4 nm to +4 nm. c. Fourier transform of the estimated phase (bottom left). d. Fourier transform of the estimated log-amplitude (bottom right).}
\label{fig:rec_ampl}
\end{figure}

Using these images, we retrieve the COFFEE estimates of the phase \(\widehat{\phiu^1}\) and the log-amplitude \(\widehat{\xiu^1}\). We then subtracted the reference phase \(\widehat{\phiu^0}\) and the reference log-amplitude \(\widehat{\xiu^0}\) and obtain \(\widehat{\phiu}\) and \(\widehat{\xiu}\), which are displayed in Fig.~\ref{fig:rec_ampl}, along with their Fourier transforms. The complete set of parameters used for the reconstruction is shown in Table~\ref{tab:param_ampl}.

\begin{table}
\caption{Parameters of reconstruction for the wavefront dominated by amplitude aberrations\LEt{please remove the italics from a priori.}.}
\tiny
\centering
\begin{tabular}{l r}
\hline
Parameter & Value\\
\hline
Wavelength \(\lambda\) & 783.25\,nm \\
Data set & \(\mathbf{I}^1\) \\
Images size & \(360\times360\) pixels \\
Sampling factor & 7.14 \\
Lyot ratio & 0.759 \\
Diversity phases & \(\phi_{\text{div},(k=1)}\) and \(\phi_{\text{div},(k=2)}\) \\
Light flux on the photometer for \(\mathbf{I}^1_{k=0}\) & \(2.5\times10^{-6}W\)\\
Light flux on the photometer for \(\mathbf{I}^1_{k=1}\) & \(5.2\times10^{-7}W\)\\
Light flux on the photometer for \(\mathbf{I}^1_{k=2}\) & \(1.5\times10^{-7}W\)\\
Number of averaged frames for \(\mathbf{I}^1_{k=0}\) & 1600 \\
Number of averaged frames for \(\mathbf{I}^1_{k=1}\) & 1100 \\
Number of averaged frames for \(\mathbf{I}^1_{k=2}\) & 1600 \\
Read-out noise standard deviation & 1 electron \\
Coronagraph type & 4-quadrant phase mask \\
A priori on \(\sigma_{\phiu^1}\) & 3\ nm\\
A priori on \(\sigma_{\phid^1}\) & 0.5\ nm\\
A priori on \(\sigma_{\xiu^1}\) & 3\ nm\\
\hline
\end{tabular}

\label{tab:param_ampl}
\end{table}

The reconstruction is visibly dominated by the introduced amplitude sinusoid: the main pair of spots is clearly visible in the log-amplitude reconstruction and does not appear at all in the phase reconstruction. The corresponding secondary pair of spots has both log-amplitude and phase components. This is not surprising: propagation of an off-pupil aberration at a frequency different from the Talbot frequency has no reason to yield only amplitude in a pupil plane. The equivalent root mean square value of the COFFEE-estimated sinusoid is \(\sigma_{\hat{\xiu}} = 1.5\) nm. The control voltage of the introduced aberration was calculated so that its root mean square value would be 1.6 nm.

\section{Experimental retrieval of a known wavefront mixing phase and amplitude aberration}
\label{sec:onde}
\begin{figure}[!ht]
\centering
\includegraphics[height=0.33\linewidth]{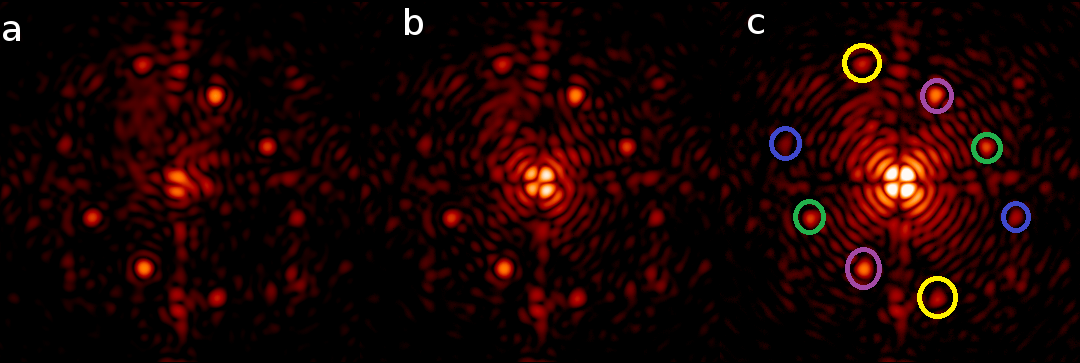}\includegraphics[height=0.33\linewidth]{colorbar}
\caption{a. Focused image \(\mathbf{I}^2_{k=0}\) (left). b. Diversity image \(\mathbf{I}^2_{k=1}\) (middle). c. Diversity image \(\mathbf{I}^2_{k=2}\) (right). The scale is an argument hyperbolic sine.  On the rightmost image, the main pair of amplitude spots is enhanced by green circles, and the corresponding replica spots are enhanced by blue circles. The main pair of phase spots is enhanced by purple circles, and the corresponding replica spots are enhanced by yellow circles. They are clearly visible in all three images.}
\label{fig:I2}
\end{figure}

During the experiment described in this Section, we kept the same sinusoid on the amplitude mirror DM1. On the phase mirror, DM3, we added a phase sinusoid of frequency \(\mu\).
The corresponding data \(\mathbf{I}^2_{k=0}\) is acquired and displayed in Fig.~\ref{fig:I2}. \LEt{the following two sentences seem to be simply descrition of the figure. Please move this information to the figure legend.}\LEt{Symbol types etc. should be described in the figure legend and not repeated in the main text. Please refer to "Tables and figures" in the author information http://www.aanda.org/author-information/paper-organization}On the rightmost image of Fig.~\ref{fig:I2}, as in Fig.~\ref{fig:I1}, the green circles show the main pair of spots due to the amplitude mirror, DM1, and the blue ones indicate replica spots of DM1. The purple circles show the main pair of spots due to the phase mirror, DM3, and the yellow one shows replica spots of DM3.

 Using these images, we retrieve the COFFEE estimates of the phase \(\widehat{\phiu^2}\) and the log-amplitude \(\widehat{\xiu^2}\). We then subtracted the reference phase \(\widehat{\phiu^0}\) and the reference log-amplitude \(\widehat{\xiu^0}\) and obtain \(\widehat{\phiu}\) and \(\widehat{\xiu}\).
 
 The complete set of parameters used for the reconstruction is shown in Table~\ref{tab:param_onde}.
We compare our estimation to the self-coherent camera (SCC) measurement that is routinely used on THD2. The SCC uses the stellar light diffracted by the FQPM outside of Lyot stop to create an additional beam (called here reference pupil). As in the Young’s experiment, the coherence between the stellar light in this beam and in the Lyot stop generates fringes in the focal plane and spatially encodes the speckles. This spatial modulation allows a direct measurement of the complex amplitude of the electric field in the recorded focal plane (\cite{galicher2008wavefront}). As shown in \cite{Mazoyer-a-13}, we are also able to retrieve the field upstream of the coronagraph using the complex field in the focal plane downstream of a phase mask coronagraph.

Using the same images than before $I^i_{k=0}$, we estimated the upstream field using Eq 28. in \cite{Mazoyer-a-13}. In this equation, the field directly measured in the focal plane by the SCC is divided by the coronagraph function $\mathcal{M}$ and the focal plane field $A_R$ of the SCC reference pupil and is normalized by the input source flux. Comparing recorded and simulated images for a given known aberration (here a sinusoid created by the phase deformable mirror) allows the normalization of the phase as a function of the intensity on the camera and the source input flux measured by the photometer. The FQPM mask $\mathcal{M}$ was assumed to be perfect. The image corresponding to the diffraction of the SCC reference pupil in the focal plane is recorded separately on the camera. An azimuthal average of this image was used to limit the impact of the noise of $A_R$. \revision{To avoid division by zero, the division was restricted to an area larger than the corrected region (\(30 \lambda/D \times 30 \lambda/D\) \revisionb{and we suppressed} the estimation of higher spatial frequencies.} 

We assumed downstream aberrations are limited to an optical path difference (OPD) between $A_R$ and the main beam and a downstream tip-tilt. As explained in §4.5.2 in \cite{Mazoyer-a-13}, we calculated the OPD which minimizes the amplitude on the complex field while introducing only phase aberration with the phase deformable mirror. The downstream tip and tilt are calculated the same way by minimizing the amplitude estimated when introducing only phase aberrations.

\begin{figure}[!ht]
\centering
\includegraphics[width=\linewidth]{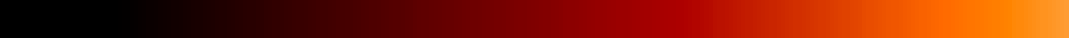} 

\vspace{-2mm}
\includegraphics[width=\linewidth]{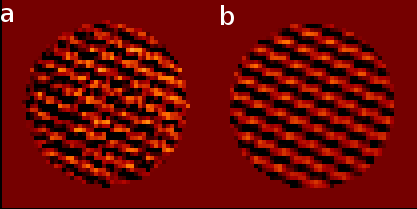}

\includegraphics[width=\linewidth]{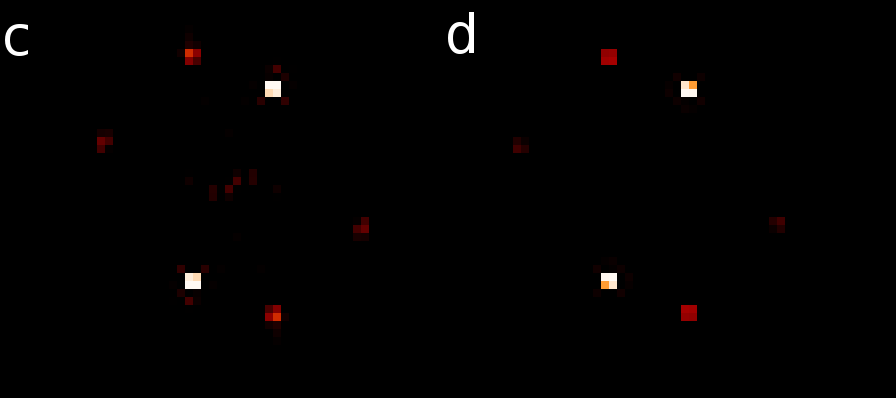}
\caption{a. Estimated phase using COFFEE (top left). b.estimated phase using the SCC (top right). The linear color bar extends from -8 nm to +8 nm. c. Fourier transform of the estimated phase using COFFEE (bottom left). d. Fourier transform of the estimated phase using the SCC (bottom right). Linear scale.}
\label{fig:rec_phase_onde}
\end{figure}

\begin{figure}[!ht]
\centering
\includegraphics[width=\linewidth]{colorbar_lineaire} 

\vspace{-2mm}
\includegraphics[width=\linewidth]{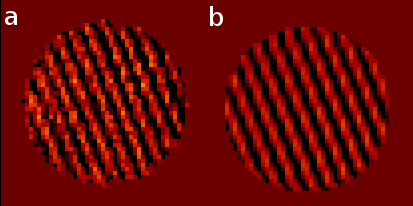}

\includegraphics[width=\linewidth]{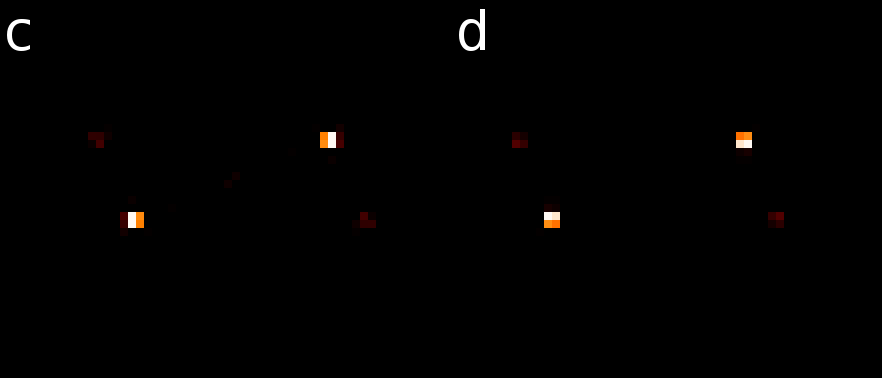}
\caption{a. Estimated log-amplitude using COFFEE (top left). b. Estimated log-amplitude using the SCC (top right). The linear color bar extends from -4 nm to +4 nm. c. Fourier transform of the estimated log-amplitude using COFFEE (bottom left). Fourier transform of the estimated log-amplitude using the SCC (bottom right). Linear scale.}
\label{fig:rec_ampl_onde}
\end{figure}

 Figure~\ref{fig:rec_phase_onde} displays the COFFEE and SCC reconstructions of \(\phiu\) along with their Fourier transforms. The correlation between the COFFEE and the SCC phase estimation is 86\%. The root mean square value of the COFFEE phase reconstruction is 3.0 nm versus 2.9 nm for the SCC one. 
  Figure~\ref{fig:rec_ampl_onde} displays the COFFEE and SCC reconstructions of \(\xiu\) along with their Fourier transforms. The correlation between the COFFEE and the SCC log-amplitude estimation is 89\%. The root mean square value of the COFFEE log-amplitude reconstruction is 1.7 nm versus 1.6 nm for the SCC one.
  Several factors contribute to the discrepancy in the correlations. While the SCC data are taken with a tip-tilt stabilization loop closed, the COFFEE data had to be taken with the tip-tilt loop open. Consequently, there is a tip-tilt phase difference between the SCC and the COFFEE estimate. Another factor is that there is a sub-pixel centering difference between the COFFEE and the SCC estimates. Finally, the COFFEE estimates and the SCC estimates are simply not identical.

\begin{table}
\caption{Parameters of reconstruction for the wavefront mixing phase and amplitude aberration\LEt{please remove the italics from a priori.}.}

\centering
\tiny
\begin{tabular}{l r}
\hline
Parameter & Value\\
\hline
Wavelength \(\lambda\) & 783.25\,nm \\
Data set & \(\mathbf{I}^2\) \\
Images size & \(360\times360\) pixels \\
Sampling factor & 7.14 \\
Lyot ratio & 0.759 \\
Diversity phases & \(\phi_{\text{div},(k=1)}\) and \(\phi_{\text{div},(k=2)}\) \\
Light flux on the photometer for \(\mathbf{I}^2_{k=0}\) & \(2.5\times10^{-6}W\)\\
Light flux on the photometer for \(\mathbf{I}^2_{k=1}\) & \(5.2\times10^{-7}W\)\\
Light flux on the photometer for \(\mathbf{I}^2_{k=2}\) & \(1.5\times10^{-7}W\)\\
Number of averaged frames for \(\mathbf{I}^2_{k=0}\) & 1200 \\
Number of averaged frames for \(\mathbf{I}^2_{k=1}\) & 800 \\
Number of averaged frames for \(\mathbf{I}^2_{k=2}\) & 1000 \\
Read-out noise standard deviation & 1 electron \\
Coronagraph type & 4-quadrant phase mask \\
A priori on \(\sigma_{\phiu^0}\) & 3\ nm\\
A priori on \(\sigma_{\phid^0}\) & 0.5\ nm\\
A priori on \(\sigma_{\xiu^0}\) & 3\ nm\\
\hline
\end{tabular}

\label{tab:param_onde}
\end{table}

The main characteristics of the phase and amplitude aberrations are retrieved. This is best visible by examining the Fourier transforms of the aberrations. 

In the Fourier transform, the main pair of spots (circled in green in Fig.~\ref{fig:I1})  generated by the amplitude mirror, DM1, is clearly visible in the log-amplitude reconstruction and does not appear at all in the phase reconstruction. The corresponding secondary spots (circled in blue in Fig.~\ref{fig:I1}) still have both log-amplitude and phase components because, as explained in Sect.~\ref{sec:ampl}, propagation of an off-pupil aberration at a frequency different from the Talbot frequency has no reason to yield only amplitude in a pupil plane.

The main pair of spots which is generated by the phase mirror, DM3, is very bright and visible in the phase reconstruction, and does not appear in the amplitude reconstruction. The corresponding secondary pair of spots appears only in the reconstructed phase but not in the reconstructed amplitude. This is expected: since the phase mirror is in a pupil plane, it has influence only on the phase and no influence on the amplitude. These characteristics of the reconstructed wavefront are proof that coronagraphic phase diversity is able to reconstruct both phase and amplitude from coronagraphic focal-plane images. 

\begin{figure}[!ht]
\centering
\includegraphics[width=\linewidth]{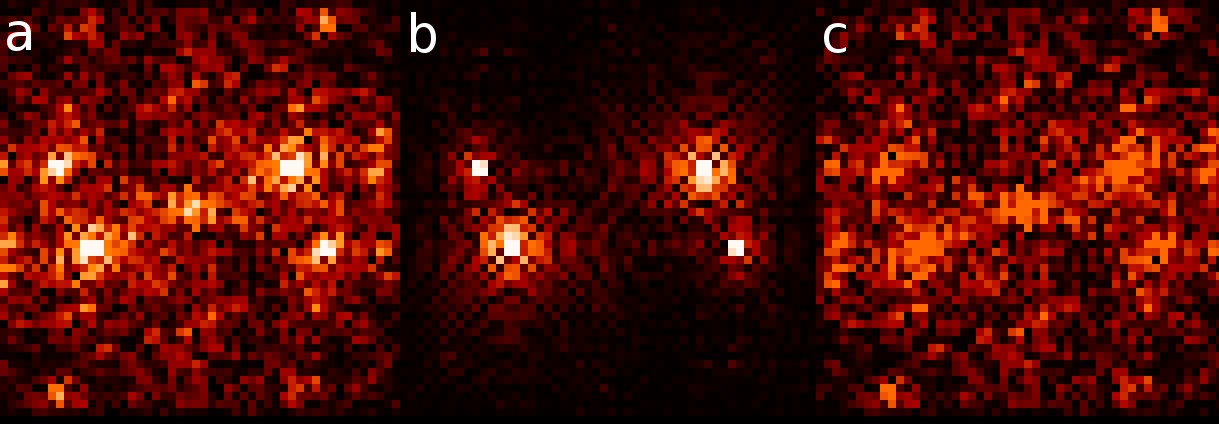}
\caption{a. COFFEE-estimated log-amplitude (1.7 nm RMS, left). b. SCC-estimated log-amplitude (1.6 nm RMS, middle). c. Absolute value of the difference (0.5 nm RMS, right). The scale is an argument hyperbolic sine.}
\label{fig:difference}
\end{figure}

\revision{Figure \ref{fig:difference} displays, on a very nonlinear scale, the differences between the COFFEE estimate and the SCC estimate in a focal plane (which are displayed on a linear scale in Fig. \ref{fig:rec_ampl_onde}). The difference between those 1.6 nm RMS estimates amounts to 0.5 nm RMS. \revisionc{Four} different contributions might explain this residual difference\LEt{A\&A avoids numbered lists. Please consider re-formatting this to plain text.}. Firstly, the two methods use different data set to perform the estimate. COFFEE uses focal and diversity images while the SCC uses fringed images. Secondly, high spatial frequencies are not estimated by the SCC, and some low frequencies might be partially unseen by the SCC. Thirdly, COFFEE may reconstruct spurious aberrations if there is a mismatch between the computer model used in the reconstruction and the actual instrument. Fourthly, despite the regularization, the noise present in the data might induce a residual noise in the COFFEE reconstruction. This could be alleviated at the cost of longer exposures or by introducing a regularization specific to the sinusoidal aberration profile that we used for the sake of the experiment.  
}

\section{Conclusion}

In this paper, we developed an extension of coronagraphic phase diversity to the estimation of the complex electric field, that is, the joint estimation of phase and amplitude. 
We demonstrate experimentally on the \textit{Très Haute Dynamique} testbed at Observatoire de Paris that coronagraphic phase diversity is able to reconstruct phase and amplitude aberrations with a subnanomtric precision. Finally, we performed the first comparison between the  complex wavefront estimated using coronagraphic phase diversity (which relies on time-modulation of the speckle pattern) and the one reconstructed by the self-coherent camera (which relies on the spatial modulation of the speckle pattern); and we found a good agreement between the two methods.
This paves the way to coronagraphic phase diversity as a coronagraphic wave-front sensor candidate for very high contrast space missions.

The next step of our work will be to use the fine knowledge of aberrations as a ground for practical implementation of the nonlinear dark hole (\cite{Paul-a-13b}), which uses a dual formalism of coronagraphic phase diversity in order to minimize the speckle intensity in the focal plane.

\begin{acknowledgements}
The PhD work of O. Herscovici-Schiller is co-funded by CNES and ONERA. 
We thank J.-M. Le Duigou (CNES) for his support. 
This work received funding from the E.U. under FP7 Grant Agreement No. 312430 OPTICON, from the CNRS (Défi Imag’In) and from ONERA in the framework of the VASCO research project.
OHS wishes to thank A. Montmerle Bonnefois for a very helpful discussion on the description of geometric aberrations, N. Védrenne for a discussion on some parametrization and numeric instabilities issues, and L. Prengère for a discussion on the domain of validity of linear filtering techniques for optical imaging.
We thank the reviewer for constructive criticism and valuable comments that helped to improve this paper. \revisionc{We thank the language editor, G. Rodgers, for corrections that helped to improve this paper.}
\end{acknowledgements}

\appendix
\section{Unseen modes of the four-quadrant phase mask coronagraph}
\label{ann:modes_non-vus}

\LEt{I\ have assumed that these sections are general methods, and left them in the present tense.}We have seen that a model of the four-quadrant phase mask coronagraph may create numerical difficulties during the reconstruction of aberrations. These troubles call for a specific regularization, such as that proposed in Eq.~(\ref{eq:regul_4Q}). Let us examine the image of an on-axis source behind a four-quadrant phase mask coronagraph~(\cite{rouan2007corono4Q}), in the spirit of Jean Gay’s analysis in \cite{abe2003phase}.
The on-axis image is given by Eq.~(\ref{eq:hc}), and here the focal-plane mask of the coronagraph writes \(\M(\alpha) = \text{Sign}(\alpha_x) \times \text{Sign}(\alpha_y) \). Let us define the two-dimensional Hilbert transform as \(\mathcal{H}=\mathcal{H}_y[\mathcal{H}_x]\), where \(\mathcal{H}_x\) and \(\mathcal{H}_y\) are the usual Hilbert transform along the first and second Cartesian coordinates,
\begin{equation}
\mathcal{H}_x[\psi](x) = \frac{1}{\pi}\dashint_\mathbb{R}\frac{\psi(x')}{x-x'}\dd x' .
\end{equation}
An equivalent formulation of Eq.~(\ref{eq:hc}) is
\begin{equation}
h_\text{c}[\psiu,\psid](\alpha) = \left|\TF^{-1}\left[ \Pd \psid \right] \star \left[ \M \TF^{-1}(\Pu\psiu) \right] \right|^2 (\alpha),
\end{equation}
where \(\star\) denotes the convolution product, \(\psiu=\exp[\ii \phiu+\xiu]\), and \(\psid=\exp[\ii\phid+\xid]\).

We can use the fact that for any function \(\psi\), \mbox{\( \TF^{-1}\{\mathcal{H}[\psi]\}(\alpha)= -\M(-\alpha)\ \TF^{-1}[\psi](\alpha)  \)} and the fact that \(\forall \alpha, \M(-\alpha) = \M(\alpha)\) to transform this expression into
\begin{equation}
h_\text{c}[\psiu,\psid](\alpha) = \left|\TF^{-1}\left[ \Pd\psid \right] \star \TF^{-1} \left[ \mathcal{H}(\Pu\psiu) \right](\alpha) \right|^2.
\end{equation}
Let us analyze the upstream complex fields \(\psiu\) such that \(h_text{c}\) is zero.
If \(\mathcal{H}(\Pu \psiu)\) is zero where \(\Pd\) is not, then \(h_\text{c}[\psiu,\psid]\) is zero everywhere. Let us analyze the nullity condition on the Hilbert transform. We note \((\alpha_x,\alpha_y)=\alpha\) the coordinates of the focal-plane position \(\alpha\) and \((r_x,r_y)=r\) the coordinates of the pupil-plane position \(r\).

\begin{align}
&\forall r, \mathcal{H}[\Pu\psiu](r) = 0 \nonumber \\
&\Leftrightarrow  \forall r, \dashint\dashint\dfrac{\Pu\psiu(r-r')}{r'_x r'_y} \dd r' =0 \\ 
& \Leftrightarrow  \forall r, \dashint \dashint \dfrac{(\Pu\psiu) \star \delta_{r'}(r)}{r'_x r'_y} \dd r' =0 \\ 
& \Leftrightarrow  \forall \alpha, \dashint \dashint \dfrac{\TF[\Pu\psiu](\alpha) \times \e^{-\ii 2 \pi r' \cdot \alpha}}{r'_x r'_y} \dd r' =0 \\ 
& \Leftrightarrow  \forall \alpha, \TF[\Pu\psiu](\alpha) \dashint \dfrac{ \e^{-\ii 2 \pi r'_x \alpha_x}}{r'_x} \dd r'_x \dashint \dfrac{ \e^{-\ii 2 \pi r'_y \alpha_y}}{r'_y} \dd r'_y =0. \label{eq:nulliteTH} 
\end{align}
The right-hand sign of the last equivalence, Eq.~(\ref{eq:nulliteTH}), dictates that 
\begin{equation}
\forall \alpha, \alpha_x \neq 0 \wedge \alpha_y \neq 0 \Rightarrow \TF[\Pu\psiu](\alpha) = 0.
\end{equation}
If \(\alpha_y = 0\) and \(\alpha_x \neq 0\), that is, if \(\alpha\) lies on the \(x\)-axis, then Eq.~(\ref{eq:nulliteTH}) reduces to
\begin{equation}
\forall \alpha, \alpha_y = 0 \Rightarrow \TF[\Pu\psiu](\alpha) \dashint \dfrac{ \e^{-\ii 2 \pi r'_x \alpha_x}}{r'_x} \dd r'_x \dashint \dfrac{1}{r'_y} \dd r'_y  =0.
\end{equation}

\begin{figure}[!ht]
\centering
\includegraphics[width=\linewidth]{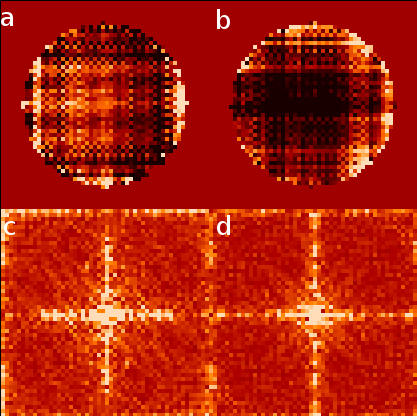}
\caption{Effect of the absence of appropriate regularization on a COFFEE reconstruction performed on the same data as in Sect.~\ref{sec:onde}. a. Estimated phase (top left). b. Estimated log-amplitude (top right). c. Fourier transform of the estimated phase (bottom left). d. Fourier transform of the log-amplitude (bottom right).}
\label{fig:sans_regul}
\end{figure}

And \(\dashint 1/r'_y \dd r'_y  =0 \), independently of \(\alpha\). Of course the same behavior happens if \(\alpha\) belongs to the \(y\)-axis.
We conclude that, mathematically, the four-quadrant phase mask coronagraph is insensitive to any upstream aberrations whose Fourier transform is nonzero only on the transitions of the four-quadrant phase mask. This condition that the Fourier transform of the aberration be infinitely thin in the focal plane implies that the aberration is of infinite extension in the pupil plane, which is physically inconsistent. 
However, as far as numeric computations are concerned, any mode whose Fourier transform is significantly different from zero only on a region of width one pixel around the axes of the four-quadrant phase mask is unseen. The result is a lack of injectivity of the model of image formation, and an adapted regularization is thus necessary. Figure~\ref{fig:sans_regul} shows the impact of the absence of regularization on a COFFEE reconstruction. The standard deviation of the estimated phase is 34.0 nm; the standard deviation of the estimated log-amplitude is 34.9 nm. Both these figures are too big by an order of magnitude, and the structures of the Fourier transforms of the estimated phase and log-amplitude are completely overwhelmed by the unseen modes of the four-quadrant phase mask coronagraph. In contrast, the reconstructions shown in Figs.~\ref{fig:rec_phase_onde} and~\ref{fig:rec_ampl_onde} with the novel regularization of Eq.~\ref{eq:regul_4Q} do not include these modes and has the expected root mean square value.  

\section{Consequences for the focal-plane in the discretization of a pupil-plane sinusoid}
\label{ann:DM}
The experiments that we present in this article rely on the use of focal-plane images obtained by imposing a sinusoid on a segmented deformable mirror. However, as we observe in Fig.~\ref{fig:I1} or even more clearly in Fig.~\ref{fig:rec_ampl}, instead of the expected pair of spots corresponding to a true sinusoid, we observe two pairs of spots (at least). The presence of the supplementary spots is due to the discrete nature of the segmented deformable mirror. In order to understand this phenomenon, we write explicitly the propagation of light from a segmented mirror in a pupil plane to a focal plane. As with the rest of the paper, we work in the framework of Fourier optics. We make the additional assumption that the mirror is square, and that the influence functions of the deformable mirror are perfect squares. Since the electric field in the focal plane in the two-dimensional case is a product of independent one-dimensional solutions, we make most of calculations in one dimension for the sake of clarity.

Let us consider a phase aberration that consists in an imaginary exponential of amplitude \(C\) and of frequency \(\nu\) that is discretized on the N different independent actuators of a square segmented mirror of side dimension \(D\).
If we denote by \(A(r)\) a discretized approximation of pupil-plane coordinate \(r\) on the mirror, the electric field in the pupil is

\begin{equation}
\mathcal{P}(r) = \exp\left\{ \ii C \sin[2 \pi \nu A(r)] \right\}.
\end{equation}

In our case, \(C \ll 1\). So we can perform a first-order Taylor expansion:
\begin{equation}
\mathcal{P}(r) \approx 1 + \ii C \sin[2 \pi \nu A(r)].
\end{equation}

The influence of the phase aberration in the pupil plane is, at first order, entirely encoded in \(\sin[2 \pi \nu A(r)]\), so we now compute the corresponding electric field in the focal plane. 
For the sake of simplicity of the calculation of this electric field, we decompose the sinus as a difference of imaginary exponentials, and calculate the electric field \(\mathcal{E}_{\nu,N}(\alpha)\) corresponding to an imaginary exponential.

\begin{align}
& \mathcal{E}_{\nu,N}(\alpha)  = \int_{-D/2}^{D/2}\exp[\ii 2 \pi \nu A(r)]\exp[-\ii 2 \pi \alpha r/\lambda] \dd r \\
 & \hspace{3cm} \text{\small (Fraunhofer propagation)} \nonumber \\
 & = \sum_{n=-N/2 + 1/2}^{N/2-1/2} \int_{nD/N-D/(2N)}^{nD/N+D/(2N)} \exp[\ii 2 \pi \nu A(r)]\\
 &\hspace{1cm}\times \exp[-\ii 2 \pi \alpha r/\lambda] \dd r \nonumber \\
  & \hspace{3cm} \text{\small (Chasles relation)} \nonumber \\
 & = \sum_{n=-N/2 + 1/2}^{N/2-1/2} \int_{nD/N-D/(2N)}^{nD/N+D/(2N)} \exp[\ii 2 \pi \nu n D/N]\\
 &\hspace{1cm}\times \exp[-\ii 2 \pi \alpha r/\lambda] \dd r \nonumber \\ 
 & \hspace{3cm} \text{\small (\textit{A} is piecewise constant)} \nonumber
\end{align}
\begin{align}
& = \sum_{n=-N/2 + 1/2}^{N/2-1/2} \exp\left[\ii 2 \pi \frac{\nu n D}{N}\right] \dfrac{\lambda}{-\ii 2 \pi \alpha } \nonumber \\
&\hspace{1cm} \times \left\{\exp\left[-\ii 2 \pi \alpha D \frac{n+1/2}{\lambda N}\right]\right.  \\
&\hspace{1.5cm}\left. - \exp\left[-\ii 2 \pi \alpha D \frac{n-1/2}{\lambda N}\right]\right\}\nonumber\\ 
 & \hspace{3cm} \text{\small (Integration over \textit{r})} \nonumber 
 \end{align}
 \begin{align}
  & = \sum_{n=-N/2 + 1/2}^{N/2-1/2} \dfrac{\lambda}{-\ii 2 \pi \alpha } \exp\left[\ii 2 \pi n \left( \nu \frac{D}{N} - \frac{\alpha D}{\lambda N} \right) \right] \\
  & \hspace{1cm} \times \left\{\exp\left[\frac{-\ii \pi \alpha D }{\lambda N}\right]-\exp\left[\frac{\ii \pi \alpha D }{\lambda N}\right]\right\} \nonumber \\
 & \hspace{3cm} \text{\small (Factorization)} \nonumber 
 \end{align}
  
 \begin{align} 
  & = \frac{D}{N} \sinc\left[ \frac{\pi \alpha D}{\lambda N} \right] \\
  & \hspace{1cm}\times \sum_{n=-N/2 + 1/2}^{N/2-1/2} \left\{\exp\left[\ii 2 \pi \left( \frac{\nu D}{N} - \frac{\alpha D}{\lambda N} \right) \right]\right\}^n \nonumber \\ 
 & \hspace{3cm} \text{\small (Definitions of sinus and cardinal sinus)} \nonumber  \\ 
 & =  \dfrac{\exp\left[ \ii \pi\frac{D}{N} (-N+1) \left( \nu - \frac{\alpha }{\lambda } \right) \right] - \exp\left[\ii \pi\frac{D}{N} (N+1) \left( \nu - \frac{\alpha }{\lambda } \right) \right]}{1-\exp\left[\ii 2 \pi \frac{D}{N} \left( \nu - \frac{\alpha }{\lambda} \right) \right]}\\
 & \hspace{1cm} \times \frac{D}{N} \sinc\left[ \frac{\pi \alpha D}{\lambda N} \right]  \qquad \text{for}\ \frac{D}{N} \left( \nu - \frac{\alpha }{\lambda} \right) \not \in \mathbb{Z} \nonumber \\ 
 & \hspace{3cm} \text{\small (Sum of a geometric series)} \nonumber 
  \end{align}
 \begin{align}
 & = \frac{D}{N} \sinc\left[ \frac{\pi \alpha D}{\lambda N} \right]\\
&  \hspace{1cm}  \times\dfrac{\exp\left[ \ii \pi D \left( \nu - \frac{\alpha }{\lambda } \right) \right] - \exp\left[-\ii \pi D \left( \nu - \frac{\alpha }{\lambda } \right) \right]}{\exp\left[\ii \pi\frac{D}{N} \left( \nu - \frac{\alpha }{\lambda } \right) \right] - \exp\left[-\ii \pi \frac{D}{N}\left( \nu - \frac{\alpha }{\lambda } \right) \right] } \nonumber  \\
 & \hspace{3cm} \text{\small (Factorization of \(\exp\left[\ii \pi \frac{D}{N} \left( \nu - \frac{\alpha }{\lambda } \right) \right]\))} \nonumber 
 \end{align}
 \begin{align}
& \mathcal{E}_{\nu,N}(\alpha)  = \frac{D}{N} \sinc\left[ \pi \frac{ \alpha D}{\lambda N} \right] \dfrac{\sin\left[ \pi D \left( \nu - \frac{\alpha }{\lambda } \right) \right]}{\sin\left[ \pi\frac{D}{N} \left( \nu - \frac{\alpha}{\lambda } \right) \right]  } \label{eq:champ_focal_sin_discret}\\
 & \hspace{3cm} \text{\small (Definition of sinus)} \nonumber
\end{align}
       
        Since a Taylor expansion shows easily that \(\lim_{x\rightarrow 0} \frac{\sin(x)}{\sin(x/N)}=N, \) the result expressed by Eq.~(\ref{eq:champ_focal_sin_discret}) is valid even if \( \left( \nu - \alpha /\lambda \right)D/N \in \mathbb{Z} \), by continuity. 
        Another Taylor expansion, this time for \(N \rightarrow \infty\) shows that
\begin{equation}
\mathcal{E}_{\nu,\infty}(\alpha) =  D \sinc\left[ \pi D \left(\nu - \frac{\alpha}{\lambda}\right) \right].
\end{equation}
Of course, this result can easily be proven by a direct calculation of the Fraunhofer propagation of the electric field with a continuous sinusoid.

An interesting property of \(\mathcal{E}_{\nu,N}\) is that, for a finite \(N\), it displays a kind of periodicity different from the usual one of \(\mathcal{E}_{\nu,\infty}\).
Indeed, for any \(\alpha\)  such that \(\alpha \neq \mp \frac{\lambda N}{D}\) 
\begin{align}
& \mathcal{E}_{\nu,N}\left(\alpha \pm \frac{\lambda N}{D}\right)  =  \\ 
& \hspace{1cm}\frac{D}{N} \sinc\left[ \pi \frac{ \alpha D}{\lambda N} \pm \pi  \right] \dfrac{\sin\left[ \pi D \left( \nu - \frac{\alpha D}{\lambda} \right) \mp \pi N \right]}{\sin\left[ \pi \left( \frac{D\nu}{N} - \frac{\alpha D}{\lambda N} \right) \mp \pi \right]  } \nonumber \\
   & \mathcal{E}_{\nu,N}\left(\alpha \pm \frac{\lambda N}{D}\right) = (-1)^N \dfrac{\frac{\alpha D}{\lambda N}}{\frac{\alpha D}{\lambda N} \pm 1} \mathcal{E}_{\nu,N}(\alpha)
\end{align}

\begin{figure}
\centering
\includegraphics[width=\linewidth]{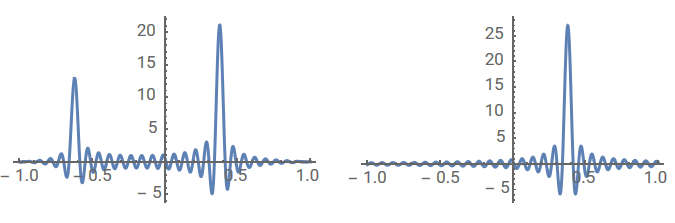}
\caption{Left: 1-dimensional focal plane electric field corresponding to a segmented approximation of a pure phase defect (\(N=32\)). Right: 1-dimensional focal plane electric field corresponding to a pure phase defect (\(N=\infty\)). The replica spot due to the approximation is clearly visible on the left figure, one correction zone left of the main spot.}
\label{fig:effet_taille_dm}
\end{figure}

This last result tells us that generating an approximate sinusoid using a segmented mirror will not only generate the expected pair of spots, but also periodic secondary pairs of spots of decreasing amplitude that would not exist if the generated phase were a true -- that is, nondiscretized -- sinusoid. Since \( \frac{\lambda N}{D} \) is the width of the frequency interval that the segmented mirror can correct, the places where these secondary spots appear in the focal plane are horizontal and vertical translations of the primary spots, with translation displacements that are multiples of the side length of the corrected zone. The closer a primary spot is to the the maximum frequency attainable by the deformable mirror (\( \frac{\pm \lambda N}{2D} \)), the closer the intensity of the first secondary spots is to the intensity of the primary spots. This effect in dimension one is displayed in Fig.~\ref{fig:effet_taille_dm}. It is two such secondary spots that are circled in blue in Fig.~\ref{fig:I1} and \ref{fig:I2}, and \revisionc{two} such secondary spots that are circled in yellow in Fig.~\ref{fig:I2}.

\bibliographystyle{aa}
\bibliography{Acronymes,biblio}   

\end{document}